\documentclass[prb,twocolumn,showpacs,showkeys]{revtex4}
\usepackage{graphicx}
\usepackage{graphics}
\usepackage{dcolumn}
\usepackage{bm}
\newcommand{\beq}{\begin{equation}}
\newcommand{\eeq}{\end{equation}}
\newcommand{\beqnar}{\begin{eqnarray}}
\newcommand{\eeqnar}{\end{eqnarray}}
\newcommand{\bfig}{\begin{figure}}
\newcommand{\efig}{\end{figure}}
\begin{document}

\title{Statistical properties of a localization-delocalization transition induced by correlated disorder}

\author{Hosein Cheraghchi$^{1,2}$, S. Mahdi Fazeli$^{1}$}

\affiliation{$^{1}$Department of Physics, Sharif University of
Technology,P.O.Box 11365-9161, Tehran, Iran \\ $^{2}$Department of
Physics, Damghan University of Basic Sciences, Damghan, Iran}
\email{cheraghchi@mehr.sharif.edu}
\date{\today}

\begin{abstract}
The exact probability distributions of the resistance, the
conductance and the transmission are calculated for the
one-dimensional Anderson model with long-range correlated
off-diagonal disorder at $E=0$. It is proved that despite of the
Anderson transition in 3D, the functional form of the
resistance (and its related variables) distribution function does not change when there
exists a Metal-Insulator transition induced by correlation
between disorders. Furthermore, we derive
analytically all statistical moments of the resistance, the
transmission and the Lyapunov Exponent. The growth rate of the
average and typical resistance decreases when the Hurst
exponent $H$ tends to its critical value ($H_{cr}=1/2$) from the insulating regime.
 In the metallic regime $H\geq1/2$, the distributions become
independent of size. Therefore, the resistance and the transmission fluctuations
do not diverge with system size in the thermodynamic limit.

\end{abstract}


 \maketitle
 \section{Introduction}
According to the pioneering work of Anderson
[\onlinecite{Anderson}], any amount of
disorder localizes all electrons in one-dimensional disordered
systems. However, in the case of off-diagonal disorder, due to chiral symmetry, there is
peculiar properties such as, divergence of the density of states and the
localization length at the band center [\onlinecite{soukoulis,cohen,ziman,cheraghchi}].
Scaling studies of the Anderson model
with diagonal disorder and a number of other models of disordered 1D systems, have found that the
dimensionless resistance $\rho=R/T$ [\onlinecite{Landauer}] satisfies $\ln(<\rho>)\propto N$ and $<\ln(1+\rho)>\propto N$
as $N$ the length of the sample goes to infinity [\onlinecite{Economou,Kramer,Andersonscale}].
Purely off-diagonal disorder at $E=0$, the same as diagonal case for $<\rho>$, shows
an exponential growth with the length,
 but in contradiction with diagonal disorder  $<\ln(1+\rho)>\propto \sqrt{N}$
[\onlinecite{Douglas,chadi}]. In the case of diagonal disorder, higher moments of the resistance
and the transmission can be calculated analytically by using
the generalized transfer matrices method [\onlinecite{Kirkman}].

In higher dimensions ($d>2$), weak disorder does not destroy the
metallic regime. Only when the strength of disorder exceeds a
critical value, the electrons become localized [\onlinecite{Anderson}]. This phenomena
which is called Anderson transition, does not depend on the
microscopic details of the system and is universal. According to the scaling theory of Abrahams,
et al [\onlinecite{Abrahams}], there exists a single parameter, conductance $g$, which determines
scaling properties of $g(N)$. Soon it became clear that the conductance $g$ is not a
self-averaged quantity. The knowledge of the mean value
$<g>$ is therefore not sufficient for complete description of
the transport properties. One has to deal with the conductance
distribution $G(g)$ [\onlinecite{Andersonscale,shapiro}] or equivalently, with
all cumulants of the conductance. This is easier in the metallic regime, where
$G(g)$ is Gaussian and the conductance fluctuations are universal [\onlinecite{Pichard}] and
independent on the value of the mean conductance and/or the system size.
The width of the distribution depends only on the dimension, the physical symmetry
of the system and depend on the boundary conditions. In the insulator,
large fluctuations of the conductance is characterized by
the log-normal distribution of $g$. Numerical studies [\onlinecite{shapiro,Markos1,Markos2}]
 proved the system-size invariance of $G(g)$ at the
critical point, which is consistent with the scaling theory of localization.
The shape of the distribution is, however, not completely understood.

In addition to the Anderson transition induced by disorder
strength, spatially correlation of disorders also can change
localization behavior of the system. A short-range spatially
correlation of the onsite disorder causes a discrete number of
extended states in the random dimer model[\onlinecite{dunlap}].
This demonstration revived interests in 1D disordered models.
Special attention has been recently paid to the presence of a
continuum of extended states in 1D around the band center in the
long-range correlated disorder model
[\onlinecite{mouraprl,izrailevprl,cheraghchi}].

Motivated by the distribution evolution of the Anderson
transition, in the present paper, we address to answer a question
about the conductance (and its related variables) distribution of
the long-range correlated off-diagonal disorder along the
metallic to insulating transition induced by the correlation
between disorders. Our calculations are done analytically at the
band center ($E=0$). It is proved that despite of the Anderson
transition in 3D induced by the disorder strength, the functional form
of the conductance distribution does not change when a phase
transition occurs by means of disorder correlation. It will be
also proved that the
conductance distribution at the phase transition point and also
in the metallic regime, has size invariance, while in 3D Anderson transition,
only the critical distribution is size independence.

Furthermore, we derive analytically all statistical moments of the transmission,
the resistance and the Lyapunov Exponent and discuss about their
fluctuations. It has been proved that for long-range correlated hopping disorder in
 the anomalously localized regime $H<1/2$, $<\rho>$ and also the typical resistance
 $\tilde{\rho}=\exp[<\ln(1+\rho)>]-1$ behaves with system size as $e^{N^{1-2H}}$
  and $e^{N^{1/2-H}}$ respectively when $N\rightarrow\infty$. Here, $H$ refers to the Hurst exponent.
  The critical Hurst exponent where occurs a localization-delocalization transition,
  was proved to be $H_{cr}=1/2$
 in this model [\onlinecite{cheraghchi}]. The growth rate of the resistance with the length
 decreases when the Hurst exponent goes to the transition point in $H_{cr}=1/2$. In the extended regime
 $H>1/2$, fluctuations of the transmission and the resistance do not diverge with system size.

This article is organized as the following sections: Section {\bf
II} introduces our model and our definition of the Lyapunov
Exponent (L.E.). The transmission distribution and its higher
moments will be calculated in Section {\bf III}. The conductance
distribution and the divergence of its moments are discussed in
Section {\bf IV}. Section {\bf V} focuses on the resistance
distribution and all its moments. Numerical results on the
distribution of the L.E. and analytical results of higher moments
at the band center are presented in Section {\bf VI}. Finally,
discussions and conclusions are presented in Section {\bf VII}.

\section{Purely Hopping Disorder Model}
We consider electrons in 1D disordered system within a tight
binding approximation. The Schroedinger equation by the nearest
neighbor assumption becomes \beq
\varepsilon_{i}\psi_{i}+t_{i,i+1}\psi_{i+1}+t_{i-1,i}\psi_{i-1}=E\psi_{i}
\eeq where E is the energy corresponding to the electron wave
function. ${\bf |\psi_{i}|}^{2}$ is the probability of finding the
electron at site i, ${\varepsilon_{i}}$ are the site potentials
which are considered here to be zero, and
${t_{i-1,i}=t_{i,i-1}=t_{i}}$ the hopping terms. Using the
transfer matrix method and the above Schroedinger equation, the
electron wave functions at the two ends of the system is related
to each other by use of total transfer matrix. The total transfer
matrix is defined as:($T_{N,0} = \prod_{i=1}^{N} T_{i,i-1}$),
where $T_{i,i-1}$ relates two wave functions at neighboring sites.
A periodic boundary condition on hopping terms as $t_{1} =
t_{N+1}$, causes to have unity determinant of total transfer
matrix for any configuration of disorder. As proved in
Ref.[\onlinecite{cheraghchi}], the L.E with the presence of the
above condition can be written in the following form.

 \beq \gamma =
\lim_{N\longrightarrow\infty}\frac{1}{N}<|{\bf F}|>_{c.a.}
\label{eq:Flyap}
 \eeq
where ${\bf F} = \ln(|{\bf a}|)$ and ${\bf a}$ is an arbitrary
eigenvalue of the total transfer matrix. $<...>$ refers to the average of the configurations.
 In the whole of this
paper, our calculations are in the special energy, ${\it E=0}$. In
this energy, the total transfer matrix can be easily derived
analytically. The function $F$ can be derived as; $ {\bf F} =
\sum_{i=1}^{k}{\bf
[}\ln(\frac{t_{2i-1}}{t_{0}})-\ln(\frac{t_{2i}}{t_{0}}){\bf ]} $
 where $\ln(t_{0}) = <\ln(t_{i})>_{c.a.}$. Randomness is imposed on the $\ln{t}$'s.
 By taking a Gaussian distribution for $\ln(t)$'s,
the above sum has a Gaussian distribution function with zero mean
for enough large system size. Therefore, the L.E of such system
has a semi-Gaussian distribution whose mean is given by:
\beq
\gamma(N) =
\sqrt{\frac{2}{\pi}}\frac{\sigma_{F}}{N}\label{eq:varF} \eeq where
$\sigma_{F}^{2}$ can be derived as a function of the pair
correlation function of $\ln(t)$'s[\onlinecite{cheraghchi}].

\begin{eqnarray}
\sigma_{F}^{2} = \frac{2}{\pi}{\bf \{}N g(0)+ 2
\sum_{\ell=1}^{N-1}(N-\ell)(-1)^{\ell}g(\ell){\bf \}}\nonumber
\\ g(i-j) = <\ln(\frac{t_{i}}{t_{0}})\ln(\frac{t_{j}}{t_{0}})>_{c.a.}\label{eq:corvar}
\end{eqnarray}
This equation is converted to the uncorrelated case with
($g(\ell) = 0$ for $\ell \neq 0$) and ($g(0) =
\sigma_{\ln(t)}^{2}$). In this case, the variance of $F$ function
would be $\sigma_{F}^{2}=N\sigma_{\ln(t)}^{2}$.
%

\section{Transmission Distribution at $E=0$}
Transmission of a particle through a one-dimensional random
potential has become a much studied problem in the theory of
disordered systems. It was shown that only the
logarithm of the transmission coefficient obeys central limit
theorem, whereas averages of $T$ and $T^{-1}$ become
unrepresentative of the ensemble for macroscopically large
systems [\onlinecite{Abrahams-Stephan}]. The logarithm of the
transmission is proportional to the L.E. and its average scales
linearly with the length of the system. Transmission fluctuations
are not damped when the sample length goes to infinity. On the other hand,
its higher moments are the same order of its mean. Finite-size
L.E. which is a self-averaging quantity in the thermodynamic
limit, is related to the transmission and conductance as the
following form.

\beq \gamma(N)=\frac{-1}{2N}\ln(T)=
\frac{1}{2N}\ln(1+\frac{1}{g}) \label{LEdef} \eeq where $g (= T /
R)$ and $T$ are conductance and transmission coefficients through
the system. Since L.E. distribution at the band center has a
semi-Gaussian distribution, so by use of the above equation,
transmission distribution $\tau(T)$ has a log-normal distribution.
 \beq
\tau(T)=\frac{1}{\sqrt{2\pi\sigma_{F}^{2}}}\frac{\exp[-\frac{(\ln
T)^{2}}{8\sigma_{F}^{2}}]}{T} \eeq
The n'th moment of transmission
can be easily extracted by using the above distribution.
 \beq
<T^{n}>=\exp({2n^{2}\sigma_{F}^{2}}) {\rm
erfc}[n\sqrt{2\sigma_{F}^{2}}] \eeq where ${\rm erfc}(x)$ is the
complementary error function, commonly denoted, is an entire
function defined by:

\beq {\rm erfc}(x)=\frac{2}{\sqrt{\pi}}\int_{x}^{\infty}e^{-t^2}dt
\eeq where for large $x$, it goes asymptotically to zero as
$\frac{e^{-x^2}}{\sqrt{\pi}x}(1-\frac{1}{2x^2}-...)$, and for
small $x$ it tends to unity as
$(1-\frac{2e^{-x^2}}{\sqrt{\pi}}(x+\frac{2x^3}{1.3}+...)$).

Depending on the disorder correlation, the variance of
($\sigma_{F}^{2}$) $F$ can be dependent or independent of the
length. In the present paper, we focus on the long-range
correlation between disorders which results in a phase transition
from the insulating to metallic phase. For long-range correlated
disorder, we consider the fluctuations of the $\ln(t)$'s are given
by the following [\onlinecite{mouraprl}]:

\beq
<[\ln(\frac{t_{i}}{t_{0}})-\ln(\frac{t_{j}}{t_{0}})]^{2}>_{c.a.}
= 2\sigma_{\ln(t)}^{2}{\bf |}\frac{i-j}{\ell_{c}}{\bf
|}^{2H}\label{eq:correlate}
 \eeq
 where $\sigma_{\ln(t)}^{2}$ is
kept fixed for all system sizes[\onlinecite{mouracom,Russ}] and
$H$ is Hurst Exponent which determines the strength of
correlation. The correlation length ($\ell_{C}$) is considered to
be equal to the system size. i,j are the positions of the bonds
along the chain. The pair correlation function arising from
Eq.(\ref{eq:correlate}), results in the following expressions for
variance of $F$.

 \beq
\lim_{N\longrightarrow\infty}\sigma_{F}^{2}(E=0)\propto\left\{\begin{array}{ll}
 \vspace{0.2 cm}
  \sigma_{\ln(t)}^{2}N^{1-2H}&H < \frac{1}{2}
\\ \vspace{0.2 cm}
\sigma_{\ln(t)}^{2}& H\geq\frac{1}{2}
\end{array}
\right.\label{eq:F-var}\eeq As shown in
Ref.[\onlinecite{cheraghchi}], an off-diagonal disordered chain
with correlation exponent of $H<\frac{1}{2}$ is anomalously
localized, while in the case of $H\geq\frac{1}{2}$, the system
has a metallic behavior. Based on the above equation, the
transmission distribution is independent of size for
$H\geq\frac{1}{2}$. Furthermore, the distribution
of all statistical variables in the metallic regime of 1D
off-diagonal disorder (such as transmission and conductance) are
independent of size, while in 3D Anderson model, only the critical distribution
is size independence. It needs to mention that in this model
 in compared with 3D Anderson model, the metallic regime is induced
 by the long-range correlation between
disorders. According to Eq.(\ref{eq:F-var}), although the width of the distributions
changes through the phase transition, but the whole shape of the distribution functions
does not change. The distributions of the conductance and the resistance will be
presented in Sections (IV,V).

Higher moments of the transmission in the anomalously
localized region, $H<\frac{1}{2}$, can be derived as
$N\rightarrow\infty$.

\beq
\lim_{N\longrightarrow\infty}<T^{n}>\sim\frac{1}{\sqrt{2\pi\sigma_{F}^{2}}}{\frac{1}{n}}\sim\frac{<T>}{n}\,\,\,;\,\,\
H<1/2
 \eeq
All moments of the transmission
 are the same order of its mean. According to the Eq.(\ref{eq:F-var}) for $H<\frac{1}{2}$,
 they go to zero with size as a power law
behavior ($N^{H-\frac{1}{2}}$). Furthermore, in the insulating regime, the fractional
variance of the transmission (the ratio of the variance to the
square of the mean) in the thermodynamic limit diverges to
infinity.

\beq \lim_{N\rightarrow\infty}{Var(T)/<T>^{2}}\rightarrow\infty
 \eeq
This behavior of the fractional variance shows that $<T>$
is not a good representative of the statistical ensemble.
Therefore the fluctuation of the transmission diverges for large samples.

A perfect transmission through the chain for large chain lengths
behaves as $\tau(T=1)\sim N^{H-\frac{1}{2}}$. The scaling behavior
of this peak is enough to explain the same behavior for the mean
value of the transmission given that other peaks of the
distribution decay in a similar or faster way with size of the
chain. In the case of uncorrelated disorder which corresponds to
$H=0$, the mean value of the transmission has an inverse square
root law in terms of the length [\onlinecite{Verges}]. This is an
interesting behavior since it perfectly coincides with scaling
predictions for wide wires of odd number of transversal modes
[\onlinecite{Mudry}].

In the metallic regime with $H\geq\frac{1}{2}$, the variance of
$F$ is independent of system size. Therefore, moments of the
transmission are independent of two variables; size and
correlation exponent. They only depend on the disorder strength.
For a weak disorder with condition $\sigma_{\ln(t)}\ll1/n$,
moments of transmission can be expanded in terms of the disorder
strength.
\beq
<T^n>_{N\rightarrow\infty}\sim1-\sqrt{\frac{8}{\pi}}n\sigma_{\ln(t)}
\,\,\,\, ; \,\,\,\, H \geq\frac{1}{2}\eeq
In the metallic regime, also it can be simply shown that the fractional variance goes to a constant in
the thermodynamic limit. Therefore, the
transmission can be a representative variable for describing
statistical properties of the 1D metallic system.

\section{Conductance Distribution}
 The distribution of the conductance can be obtained
by using its relation with L.E. in Eq.[\ref{LEdef}].
 \beq
G(g)=\frac{1}{\sqrt{2\pi\sigma_{F}^{2}}} \frac{\exp[{-\frac{{
\ln(1+\frac{1}{g})}^2 } {8\sigma_{F}^{2}}}] } {g(1+g)} \eeq

This distribution converges to a power law form
$1/(g^{2}\sigma_{F})$, as it should be for large $g$, but for
small $g$ the distribution behaves log-normal form as:
$G(g)\propto \exp[-\frac{ \ln(g)^{2} } {8\sigma_{F}^{2}}
]/{(g\sigma_{F})}$. In the metallic regime, where
$H\ge\frac{1}{2}$, variance of $F$ will be independent of the
system size. So, the conductance distribution function will be
invariant as $N\rightarrow\infty$. In the insulating regime,
$H<\frac{1}{2}$, Eq.(\ref{eq:F-var}) shows that $\sigma_F$
behaves with size as $N^{1/2-H}$ in the thermodynamic limit.
Therefore, in this regime, the conductance distribution for large
$g$, goes off more faster than the metallic one. It tends to zero
as $N^{H-1/2}/g^2$. It means that in the insulator regime, large
conductance occurrence happen rarely with a low probability. For
$g\rightarrow0$, the conductance distribution goes to zero
$G(g)\rightarrow0$. However, two limits of $g\rightarrow0$ and
$N\rightarrow\infty$ do not commute with each other.

Because the conductance distribution decays very slow
$G(g)\propto g^{-2}$ as $g\rightarrow\infty$, all moments of $g$
diverge. Large conductance can cause to diverge the mean of $g$
in the averaging process, although occurrence probability of
large conductance is very low. Therefore, for omitting the effect
of large conductance in the mean process, it is better to mean the
resistance $\rho=1/g$ instead of the conductance.

\section{Resistance Distribution}
The resistance distribution at the band center can be also derived
by using the definition in Eq.(\ref{LEdef}).

\beq
R(\rho)=\frac{1}{\sqrt{2\pi\sigma_F^2}}\frac{\exp[-\frac{[\ln(1+\rho)]^2}{8\sigma_F^2}]}{1+\rho}
\eeq This distribution goes to $(1-\rho)/\sqrt{2\pi\sigma^2_F}$
as it should be for small resistance $\rho$. So, $R(\rho)$ is an
ascending function of $\rho$ when $\rho\rightarrow0$. But for
large $\rho$, this distribution converges to the log-normal form.
The n'th moment of the resistance at $E=0$ can be derived as the
following summation.

\beq
<\rho^n>=\sum_{\ell=0}^n\frac{(-1)^{\ell}n!}{(n-\ell)!\ell!}e^{2(n-\ell)^2\sigma_F^2}{\rm
erfc}[-\sqrt{2(n-\ell)^2\sigma_F^2}]
 \eeq
 In the anomalously localized regime ($H<1/2$), where $\sigma_F^2$ goes to infinity in
 the thermodynamic limit, the above equation can be summarized as
 the following form.
 \beq
<\rho^n>=(-1)^n+2\sum_{\ell=0}^{n-1}\frac{(-1)^{\ell}n!}{(n-\ell)!\ell!}e^{2(n-\ell)^2\sigma_F^2}
 \eeq
As an immediate consequence of the above general formula, the average resistance
 grows with system size as:
\beq
<\rho>\propto e^{2 \sigma_{\ln(t)}^2 N^{1-2H}}
\eeq
Since $\frac{<\rho^n>}{<\rho>^n}\sim e^{2n(n-1)\sigma_F^2}$, the
fractional variance (n=2) of the resistance diverges as
$N\rightarrow\infty$ for $H<1/2$. Furthermore, all moments of the resistance
distribution also diverges. According to the results of Anderson
{\it et al} [\onlinecite{Andersonscale}], instead of the
resistance whose moments diverge, the appropriate quantity is a
self-averaged variable $\ln(1+\rho)(=2N\gamma)$ which its
fractional variance goes to zero as $N\rightarrow\infty$.
However, the fractional variance of the new variable in the
hopping disorder model at $E=0$ tends to a non-zero constant ($\pi/2-1$).
This constant can be derived by using a relation between the
second moment and mean of the L.E. as
$<\gamma^2>=\pi/2<\gamma>^2$. Furthermore, the average $\ln(1+\rho)$ has already calculated
 through the Eqs.(\ref{eq:varF},\ref{eq:F-var}) as:
 \beq
<\ln(1+\rho)>\propto \left\{ \begin{array}{c} \sigma_{\ln(t)} \,\,\,\,\,\,\,\,\,\,\,\,\,\,\,\,\,\ H \geq \frac{1}{2} \\
 \\ \sigma_{\ln(t)}N^{1/2-H}  \,\,\,\,\,\,\,\,\,\,\ H <
\frac{1}{2}\end{array}\right.
\eeq
The typical resistance is introduced as a candidate variable for experimental resistance.
 Using the above equation, in the thermodynamic limit, this measurable quantity
 behaves as $\tilde{\rho}\propto e^{N^{1/2-H}}$ with the system size for $H<1/2$.
 In the special case of $H=0$ which corresponds
 to the uncorrelated disorder, in good agreement with the results of Refs.[\onlinecite{Douglas,chadi}],
  the average and typical resistances are as $e^{N}$ and $e^{\sqrt{N}}$. It is clear that the growth
  rate of the average and typical resistance decreases when the Hurst exponent tends to
  its critical value $H_{cr}=1/2$.

  In the metallic regime $H>1/2$, the resistance only grows with the disorder strength. Also,
  in this regime, the fractional variance of the resistance is independent of size and also
  the Hurst exponent. This expression means that fluctuations of the resistance and the transmission
  do not increase with size when $N\rightarrow\infty$. Therefore, despite of the insulating regime,
  they can be considered as representative variables for the statistical properties of transport.

\section{Lyapunov Exponent Distribution}
 \bfig
 \begin{center}
\includegraphics[width=8 cm]{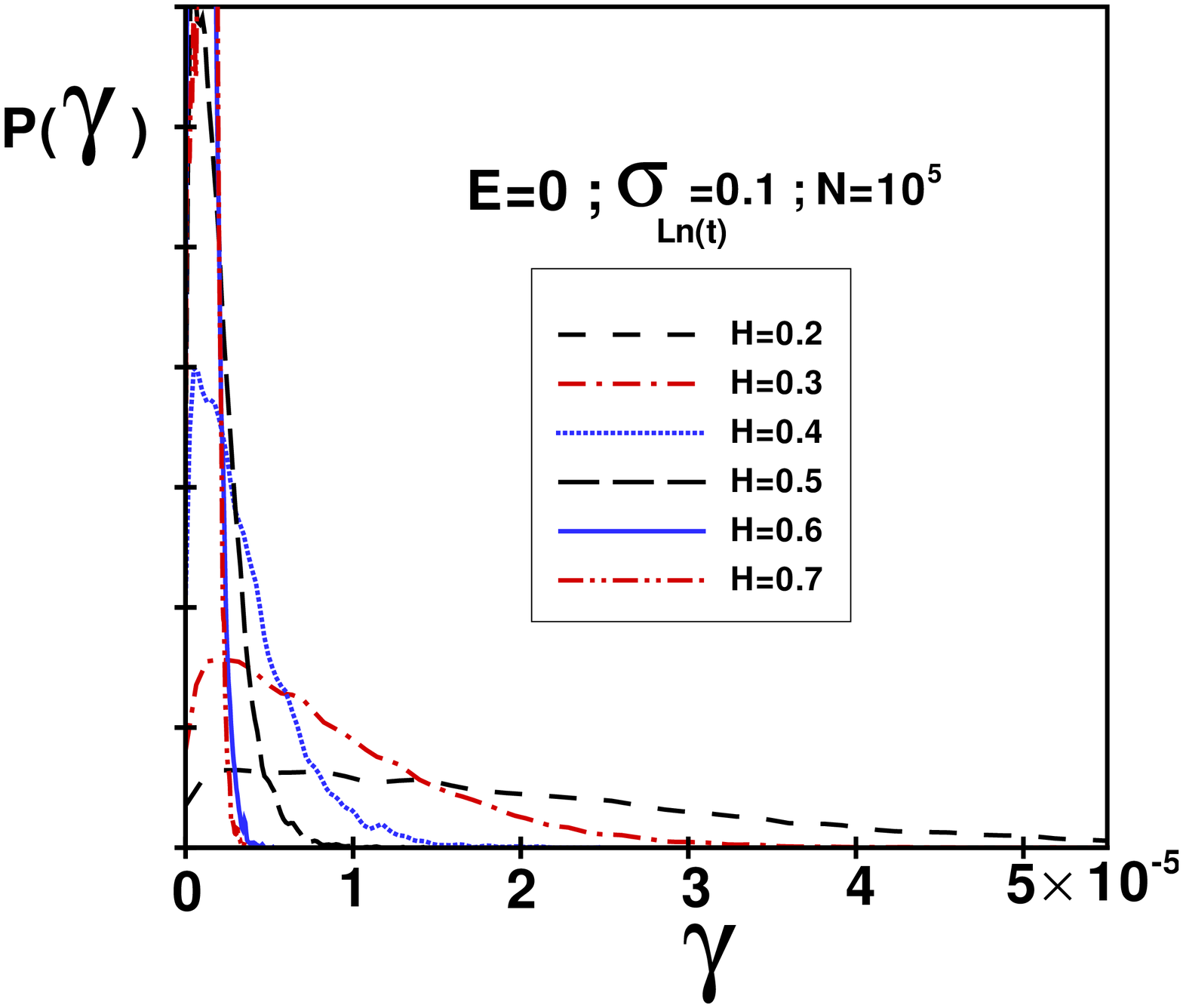}
\end{center}
\caption{Lyapunov Exponent Distribution Function for different
Hurst Exponents at $E=0$. The disorder strength is considered to be 0.1. The
sample size is $10^5$. \label{fig:distribution}}
 \efig
With care of our discussions about the transmission, the
conductance and the resistance properties, the appropriate
variable for describing localization properties is the Lyapunov
Exponent. It can be shown that the L.E. distribution function and
all its higher moments converge for large system sizes.

Fig.(\ref{fig:distribution}) shows numerical calculations of the
L.E. distribution for different Hurst Exponents
[\onlinecite{cheraghchi}]. All distribution curves have been
softened by the Kernel smoothing method [\onlinecite{Kernel}]
without changing any statistical characteristic of distributions.
It can be seen that the functional shape of the L.E. distribution does
not change when a phase transition occurs by the correlated
disorder. The pair correlation function arising from
Eq.(\ref{eq:correlate}), results in the following expression for
the L.E variance as $N\rightarrow\infty$. \beq
\sigma^{2}_{\gamma} \propto \left\{ \begin{array}{c} \frac{\sigma_{ln(t)}^{2}}{N^{2}} \,\,\,\,\,\,\,\,\,\,\,\,\,\,\,\,\,\ H \geq \frac{1}{2} \\
 \\ \frac{\sigma_{ln(t)}^{2}}{N^{1+2H}}  \,\,\,\,\,\,\,\,\,\,\ H <
\frac{1}{2}\end{array}\right. {\label{eq:landacor}} \eeq This
equation shows that correlated disorder causes the variance of
the L.E. distribution function to converge faster than in the
uncorrelated case (H=0). It can be seen that in good agreement
with Anderson {\it et al.} [\onlinecite{Andersonscale}], the
variance of the L.E. in the uncorrelated disorder scales
according to the law of large numbers as $1/N$.

With care of the semi-Gaussian distribution function for the L.E,
it can be simply proved that the n'th moment of the L.E is
proportional to the n'th power of its mean as
$<(\gamma-<\gamma>)^{n}>\propto<\gamma>^{n}$. So,
Eq.(\ref{eq:landacor}) can be generalized to higher moments of the
L.E. as the following scaling law.
 \beq
  <(\gamma-<\gamma>)^{n}>\propto \left\{\begin{array}{ll}
 (\frac{\sigma_{\ln(t)}}{N}^n& H \geq 1/2\\
\\(\frac{\sigma_{\ln(t)}}{N^{1/2+H}})^n& H < 1/2\end{array} \right.
\label{eq:cr-energy}
 \eeq
So, all moments of the L.E. in the metallic regime ($H>1/2$) are
independent of the Hurst Exponent. Fig.(\ref{fig:distribution})
shows that the variance of L.E. does not change for $H>1/2$. Of
course, at the phase transition point ($H=1/2$), the width of the
distribution will tend to $H>1/2$ values in the thermodynamic
limit.
\section{Conclusion and discussion}
The distribution functions of the resistance, the conductance and
the transmission was analytically derived for 1D off-diagonal
Anderson model with long-range correlated disorder at the band
center (E=0). Despite of the Metal-Insulator transition induced
by the disorder strength in 3D, the phase transition induced by
the correlated disorder in 1D does not cause to change the shape of the distribution
function of the conductance (and all related variables). In the
metallic regime, the distribution functions does not depend on
the system size, while in 3D Anderson model only the critical distribution
 is size independence.

We derived analytically all statistical moments of the resistance,
the transmission and the Lyapunov Exponent. In the anomalously localized regime
$H<1/2$, by means of these moments, it was shown that the fluctuations of the transmission, the
conductance and the resistance diverge in the thermodynamic limit.
Convergence of the Lyapunov Exponent will happen not only with
the system size but with the correlation exponent. The growth rate of the average and typical
 resistances decrease when the correlation exponent closes to its transition point
 from insulating to metallic regime. In the metallic regime $H>1/2$, the resistance and the transmission
 fluctuations do not diverge with the system size.
\begin{acknowledgments}
 We wish to acknowledge Prof. Keivan Esfarjani for his useful
discussion. Also, I would like to thank Prof. Peter Markos for a critical reading of the
manuscript and his effective comments.
\end{acknowledgments}

\end{document}